\documentclass[twocolumn,apsrev4,amsmath]{revtex4-1} 

\usepackage{graphicx}

\begin{document}
\title{Electrically Detected Double Electron-Electron Resonance:\\
Exchange Interaction of $^{\text{31}}$P Donors and P$_{\text{b0}}$ Defects at the Si/SiO$_{\text{2}}$ Interface}

\author{Max Suckert}
\email{max.suckert@wsi.tum.de}
\affiliation{Walter Schottky Institut, Technische Universit\"{a}t M\"{u}nchen, Am Coulombwall\,4, 85748 Garching, Germany}
\author{Felix Hoehne}
\affiliation{Walter Schottky Institut, Technische Universit\"{a}t M\"{u}nchen, Am Coulombwall\,4, 85748 Garching, Germany}
\author{Lukas Dreher}
\affiliation{Walter Schottky Institut, Technische Universit\"{a}t M\"{u}nchen, Am Coulombwall\,4, 85748 Garching, Germany}
\author{Markus Kuenzl}
\affiliation{Walter Schottky Institut, Technische Universit\"{a}t M\"{u}nchen, Am Coulombwall\,4, 85748 Garching, Germany}
\author{Hans Huebl}
\affiliation{Walther-Mei\ss ner-Institut, Bayerische Akademie der Wissenschaften, Walther-Mei\ss ner-Str.\,8, 85748 Garching, Germany}
\author{Martin Stutzmann}
\affiliation{Walter Schottky Institut, Technische Universit\"{a}t M\"{u}nchen, Am Coulombwall\,4, 85748 Garching, Germany}
\author{Martin S.~Brandt}
\affiliation{Walter Schottky Institut, Technische Universit\"{a}t M\"{u}nchen, Am Coulombwall\,4, 85748 Garching, Germany}

\date{\today}

\begin{abstract}
	We study the coupling of P$_{\text{b0}}$ dangling bond defects at the Si/SiO$_2$ interface and $^{31}$P donors in an epitaxial layer directly underneath using electrically detected double electron-electron resonance (EDDEER).
	An exponential decay of the EDDEER signal is observed, which is attributed to a broad distribution of exchange coupling strengths $J/2\pi$ from $25$\,kHz to $3$\,MHz.
	Comparison of the experimental data with a numerical simulation of the exchange coupling shows that this range of coupling strengths corresponds to $^{31}$P-P$_{\text{b0}}$ distances ranging from 14\,nm to 20\,nm.
	
\end{abstract}

\maketitle

\section{Introduction}
Design and modeling of semiconductor devices requires the detailed understanding of those defects which influence the electronic properties of these devices.
Electron paramagnetic resonance (EPR) is particularly suited to investigate the microscopic structure of paramagnetic defects in semiconductors~\cite{Spa03}. 
In samples and devices, where the number of defects is not sufficient for EPR detection, electrically detected magnetic resonance (EDMR) has been established as a versatile alternative due to its orders of magnitude higher sensitivity compared with conventional EPR~\cite{Lep72,Bra04,Mc06,Bak12,Sch12}.
Most EDMR processes involve the formation of spin pairs whose spin symmetry determines the transport properties, resulting in a resonant current change when spins are flipped by microwave irradiation~\cite{Sch66,Kap78,Kis81}.
In particular, spin-dependent recombination processes in silicon have been interpreted successfully in terms of spin pair or donor-acceptor recombination models~\cite{Sti95}.
An example of such a process is observed in phosphorus-doped silicon near the Si/SiO$_2$ interface where a spin-dependent recombination process occurs via spin pairs formed by $^{31}$P donors and P$_{\text{b0}}$ dangling bond defects~\cite{Hoe10}.
This spin pair not only serves as a prototype example for other spin-dependent recombination processes, but is also of interest in donor-based quantum information processing~\cite{Kan98,Mor10} with respect to readout of qubit states~\cite{Boe02}, providing a spin-to-charge transfer, and decoherence introduced by interface defects~\cite{Sou07}.
The advent of pulsed electrically detected magnetic resonance (pEDMR) techniques allows to address spin dynamics such as the formation and recombination of spin pairs, spin coherence, coherent control and readout of spins as well as spin-spin coupling~\cite{Ste06,Hue08,Dre12}.

Spin pairs are characterized by a noticeable coupling between the two spins due to their spatial proximity~\cite{Lu11}.
However, the coupling strength and its dependence on the spin-spin distance of such spin pairs has not been studied in detail so far.
Here, we apply electrically detected double electron-electron resonance (EDDEER) to measure the coupling between $^{31}$P donors and P$_{\text{b0}}$ Si/SiO$_2$ interface defects.
We compare the experimental results with a numerical calculation of the exchange coupling and find that the observed EDDEER signal can be attributed to a distribution of spin pairs relevant for the spin-to-charge detection scheme used here with distances between about 14\,nm and 20\,nm.

\section{Experimental methods and details}
\begin{figure}[tb]
\begin{center}
\includegraphics{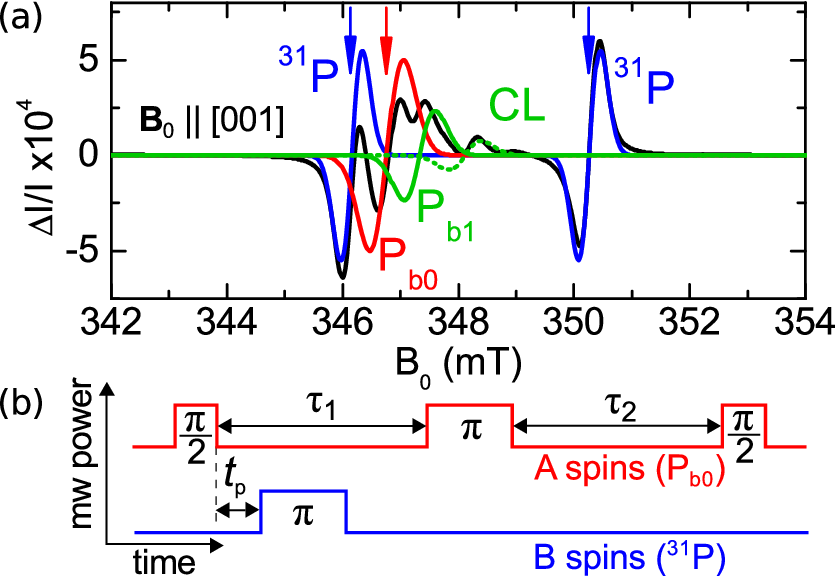}
\end{center}
\caption{\label{fig:pulseSequence}
(a)~First derivative spectrum obtained by a cwEDMR experiment~\cite{Lu11} on the Si:P sample studied here, showing the resonant change $\Delta I$ of the photocurrent $I$ (black line).
The colored curves represent a fit using Gaussian line shapes.
Hyperfine-split resonance lines of $^{31}$P with equal amplitudes of both $I = \pm 1/2$ resonances, resonance lines of the dangling bond defects P$_{\text{b0}}$ and P$_{\text{b1}}$ and a central line (CL) of exchanged coupled $^{31}$P or conduction band electrons are observed.
(b)~The basic pulse sequence of electrically detected DEER consists of a spin echo pulse sequence $\pi/2$-$\tau_1$-$\pi$-$\tau_2$-$\pi/2$ including a final $\pi/2$ projection pulse applied to the A spins plus an inversion $\pi$ pulse applied on the B spins at a time $t_{\text{p}}$ after the first $\pi/2$ pulse.
}
\end{figure}

Before we present our experimental results, we shortly review the basic idea of the DEER method, which is widely used in pulsed EPR e.g. to study the structure of complex molecules~\cite{Mil81,Jes02}. The coupling between two spins A and B is measured by the EDDEER pulse sequence sketched in Figure~\ref{fig:pulseSequence}\,(b). It consists of a $\pi/2$-$\tau_1$-$\pi$-$\tau_2$-$\pi/2$ spin echo sequence~\cite{Jes01}, including a final $\pi/2$ projection pulse~\cite{Hue08}, resonant with the A spins, where $\pi/2$ and $\pi$ denote microwave pulses with corresponding flipping angles and $\tau_1$ and $\tau_2$ periods of free evolution. At time $t_{\text{p}}$ after the first $\pi/2$ pulse an additional $\pi$ pulse is applied to invert the B spins. Taking only into account the electron spin Zeeman interaction and the exchange coupling, the spin Hamiltonian of the spin pair is given by
\begin{equation}
	\hat{\mathcal{H}} = \frac{\hbar}{2}\left(\omega_{\text{A}}\hat{\sigma}_z^{\text{A}}+\omega_{\text{B}}\hat{\sigma}_z^{\text{B}}+\frac{1}{2} J\hat{\sigma}_z^{\text{A}}\hat{\sigma}_z^{\text{B}}\right),
\label{eq:DEERhamiltonian}
\end{equation}     
where $\omega_{\text{A,B}}$ denote the Larmor frequencies of spins A and B, $J$ denotes the exchange coupling constant and $\hat{\sigma}_z$ the Pauli spin operator.
We neglect non-secular terms of the exchange coupling Hamiltonian since $\left|\omega_{\text{A}}-\omega_{\text{B}}\right| \gg J$ in this work~\cite{Jes01} as well as the dipolar coupling $D$ since it is shown below that $D \ll J$ for the spin pairs in our sample.
The last term in the Hamiltonian has the form of an additional effective magnetic field with its polarity depending on the mutual orientation of spins A and B.
Flipping spin B after a time $t_{\text{p}}$ therefore changes the local field seen by spin A, resulting in an additional phase $\Delta \phi=J \cdot t_{\text{p}}$ acquired by spin A during the spin echo pulse sequence.
This results in an oscillation of the spin echo amplitude as a function of $t_{\text{p}}$ with a frequency determined by the exchange coupling $J$. For a spin ensemble with a broad distribution of coupling constants the different oscillation frequencies will lead to a strong dephasing, eventually resulting in a decay without oscillations~\cite{Kla62}. 

%
%

In this work we use a $\sim$20\,nm-thick Si:P epilayer with a nominal P concentration of $9\times10^{16}$\,cm$^{-3}$ covered with a natural oxide and grown on a nominally undoped Si buffer on a silicon-on-insulator substrate. 
The dimensions of the sample and Ti/Au contacts for electrical measurements are the same as in Reference~\cite{Lu11}.
In this type of sample, EDMR signals originate predominantly from the $^{31}$P-P$_{\text{b0}}$ recombination process~\cite{Hoe10}. 
The sample was mounted with the silicon [001] axis parallel to the static magnetic field $\boldsymbol{B}_0$, cooled to a temperature of $\sim$5\,K, biased with 300\,mV and illuminated continuously via a glass fiber with the white light from a tungsten lamp at an intensity $\sim$30\,mW/cm$^2$, resulting in a photocurrent of 116\,$\mu$A.
We applied microwave pulses at $X$-band frequencies and adjusted the microwave power such that the $\pi$ pulse length was 30\,ns for both the $^{31}$P and P$_{\text{b0}}$ electron spins, corresponding to a microwave magnetic field $B_1$ of 0.6\,mT. 
The orientation of the sample was chosen such that the P$_{\text{b0}}$ dangling bond resonance lines are degenerate, thereby facilitating the application of microwave pulses to the P$_{\text{b0}}$.
The current transients after the pulse sequence are filtered, amplified and box-car integrated from 3\,$\mu$s to 40\,$\mu$s, yielding a charge $\Delta Q$ proportional to the amount of antiparallel spins at the end of the pulse sequence~\cite{Boe03}.
A lock-in detection scheme with a modulation frequency of $\sim$500\,Hz, corresponding to a shot-repetition time of 2\,ms, is used by applying a two-step phase cycle to the last $\pi$/2 pulse of the spin echo~\cite{Hoe12} in order to remove the background resulting from non-resonant photocurrent transients and to decrease the noise level in our measurements. 

%
\begin{figure}[htb]
\begin{center}
\includegraphics{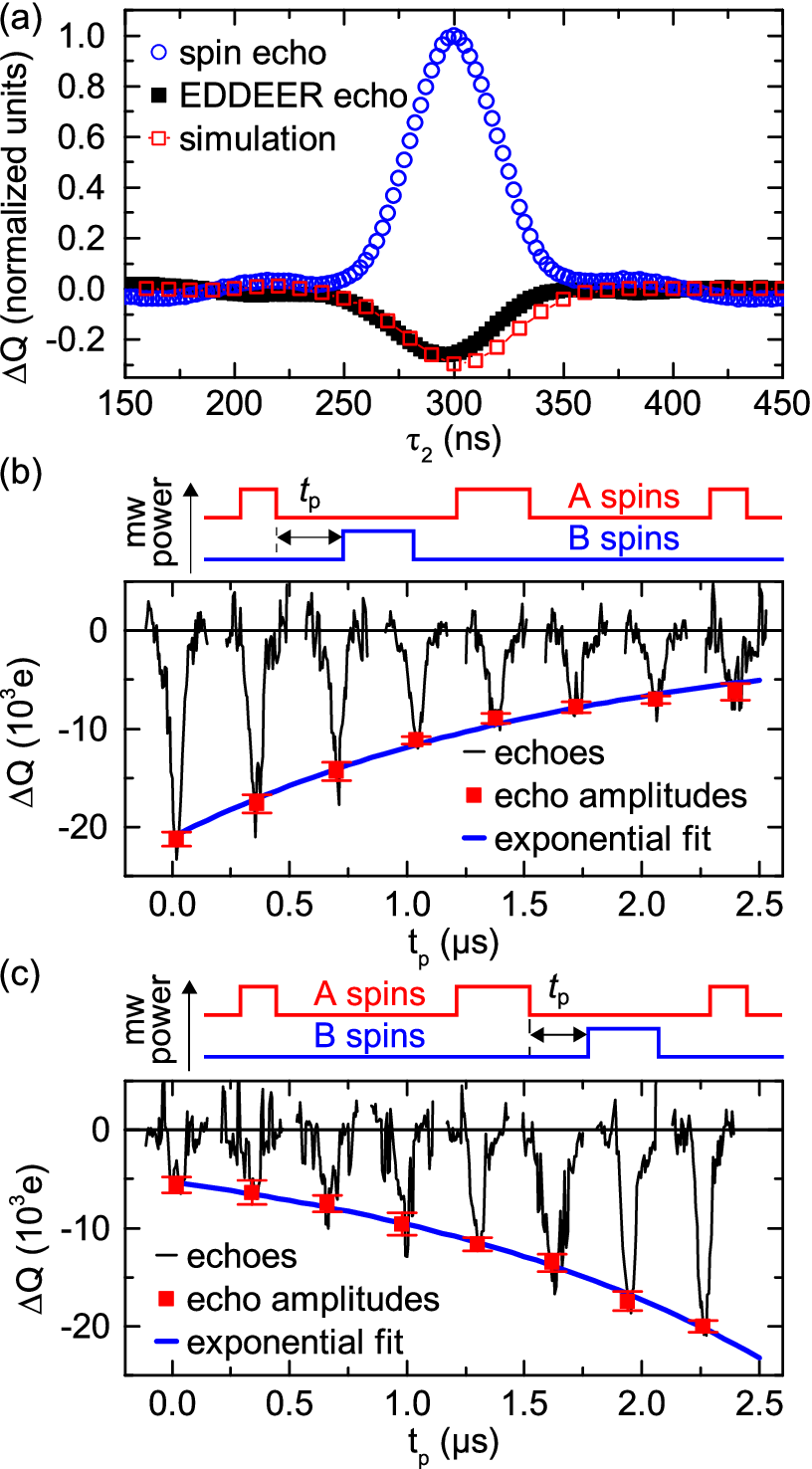}
\end{center}
\caption{
\label{fig:deerExp}
(a)~EDDEER echo recorded on the P$_{\text{b0}}$ spins for $\tau_1 = 300$\,ns and $t_{\text{p}} = 20$\,ns together with a numerical simulation of the EDDEER echo normalized to the amplitude of the corresponding spin echo. For comparison, the measured spin echo without inversion pulse on the $^{31}$P spin is shown as well.
(b)~EDDEER echoes with the inversion pulse in the free evolution interval $\tau_1 = 2.5\,\mu$s recorded for different $t_{\text{p}}$ after subtraction of a linear background. 
The corresponding amplitudes, determined by a Gaussian fit, decay with a time constant of $T$=1750$\pm$100\,ns.
(c)~EDDEER echoes for the inversion pulse in the interval $\tau_2$. Their amplitudes increase with a time constant of 1700$\pm$50\,ns.  
}
\end{figure}
%
For the EDDEER measurements, the spin echo resonantly excites the P$_{\text{b0}}$ spins [spins A, $g = 2.006$~\cite{Ste98}, indicated by the red arrow in the continuous wave (cw)EDMR spectrum shown in Figure~\ref{fig:pulseSequence}\,(a)] while the inversion pulse resonantly excites both hyperfine lines of the $^{31}$P spins [spins B, $g=1.9985$, hyperfine splitting $\Delta f_{\text{hf}} = 117.5$\,MHz~\cite{Feh59}, blue arrows in Figure~\ref{fig:pulseSequence}\,(a)].
We chose to invert the $^{31}$P spins rather than the P$_{\text{b0}}$ spins, since the smaller inhomogeneous broadening of the $^{31}$P transitions ($\text{FWHM} = 10$\,MHz) when compared to the P$_{\text{b0}}$ transition ($\text{FWHM} = 17$\,MHz) allows to invert a larger fraction of the $^{31}$P ensemble, resulting in a larger DEER signal~\cite{Lu11}.

\section{Results}
Figure \ref{fig:deerExp} (a) shows an EDDEER echo (black full squares) as a function of $\tau_2$ for $\tau_1 = 300$\,ns and $t_{\text{p}} = 20$\,ns together with a spin echo without the inversion pulse on the $^{31}$P spins (blue circles).
The EDDEER echo is inverted when compared to the simple spin echo as expected since the additional $\pi$ pulse changes the spin symmetry of the $^{31}$P-P$_{\text{b0}}$ spin pair which governs the spin-dependent recombination process.
The amplitude of the EDDEER echo signal is reduced by a factor of about four when compared to the spin echo.
This is a result of the spectral overlap of the low-field $^{31}$P and the P$_{\text{b0}}$ line as shown in Figure \ref{fig:pulseSequence}\,(a), which does not allow a fully selective excitation of the $^{31}$P spins.   
As a consequence, some P$_{\text{b0}}$ spins also are flipped by the inversion pulse on the $^{31}$P spins leading to a reduction of the echo amplitude.
This is confirmed quantitatively by a numerical simulation of the EDDEER and spin echo pulse sequences taking into account the inhomogeneous broadening of the $^{31}$P and P$_{\text{b0}}$ lines.
The result of the simulation, shown as red open squares in Figure \ref{fig:deerExp}\,(a), yields an EDDEER echo amplitude of $-0.3$ compared to the spin echo in very good quantitative agreement with the experiment.

To quantify the magnitude of the exchange coupling between the spins of the P$_{\text{b0}}$ and $^{31}$P spin pairs at the Si/SiO$_2$ interface, we recorded a series of EDDEER echoes for different time intervals $t_{\text{p}}$ with fixed $\tau_1 = 2.5\,\mu$s.
Figure \ref{fig:deerExp}\,(b) shows the echo traces, recorded as a function of $\tau_2$, centered around the respective $t_{\text{p}}$ after subtraction of a linear background.
Their amplitudes (red squares) are determined by a Gaussian fit.
The echo amplitudes decay exponentially as a function of $t_{\text{p}}$ with a decay time constant $T = 1750\pm100$\,ns, corresponding to a coupling strength of $100$\,kHz.
We interpret this decay as a result of a broad distribution of exchange couplings within the ensemble of $^{31}$P-P$_{\text{b0}}$ spin pairs which results from a distribution of $^{31}$P-P$_{\text{b0}}$ distances as discussed in detail below.

To exclude that the observed decay is indirectly caused by a recombination process with a time constant in the range of several $\mu$s~\cite{Dre12}, EDDEER echoes with the inversion pulse in the second free evolution period $\tau_2$ at $t_{\text{p}}$ after the $\pi$ pulse were recorded as shown in Figure \ref{fig:deerExp}\,(c). 
In this case, the phase acquired by spin A is given by $\Delta \phi = J(\tau_1-t_{\text{p}})$. 
For a distribution of couplings, we therefore expect an increase of the spin echo amplitude as a function of $t_{\text{p}}$ while for a recombination process as the origin of the decay in Fig.~\ref{fig:deerExp}\,(b) also a decrease would be expected here. 
As shown in Figure~\ref{fig:deerExp}\,(c), for this pulse sequence the EDDEER echo amplitude increases exponentially with a time constant $T = 1700\pm50$\,ns.
The comparison to the time-dependence of the EDDEER echo amplitude in Figure~\ref{fig:deerExp}\,(b) reveals a symmetric behavior of the EDDEER for the inversion pulses in the waiting intervals $\tau_1$ and $\tau_2$ with the same time constant within experimental uncertainty.
We therefore conclude that indeed a coupling between the P$_{\text{b0}}$ and $^{31}$P spins leads to the observed behavior rather than a recombination process.

%
\begin{figure}[tb]
\begin{center}
\includegraphics{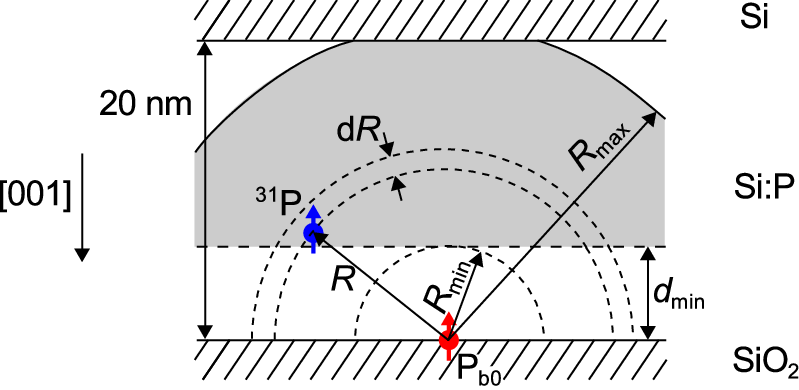}
\end{center}
\caption{\label{fig:integrationGeometry}
	Geometry used for the simulation of the observed EDDEER echo amplitude decays, based on a 20\,nm thick layer of $^{31}$P doped silicon on a Si substrate (top) and covered with a native oxide (bottom). The shaded area represents the integration range with a lower boundary $d_{\text{min}}$ and an upper boundary $R_{\text{max}}$.
	If $R_{\text{max}}$ exceeds the width of the doped layer, the spherical shell is cut as indicated.
}
\end{figure}
%
In the following we show that the time constant $T$ of the decay is a result of a distribution of $^{31}$P-P$_{\text{b0}}$ distances compatible with the width of the doped epilayer.
To this end, we numerically calculate the exchange coupling as a function of the $^{31}$P-P$_{\text{b0}}$ distance $R$. 
We further estimate the expected EDDEER decay by suitably averaging over the distribution of distances of the $^{31}$P-P$_{\text{b0}}$ ensemble. 
The geometry of the problem from the point of view of a dangling bond defect center P$_{\text{b0}}$ at the Si/SiO$_2$ interface is shown in Figure \ref{fig:integrationGeometry}.
Phosphorus donors are located in the $20$\,nm thick doped epilayer above the Si/SiO$_2$ interface.
%
\begin{figure}[htb]
\begin{center}
\includegraphics{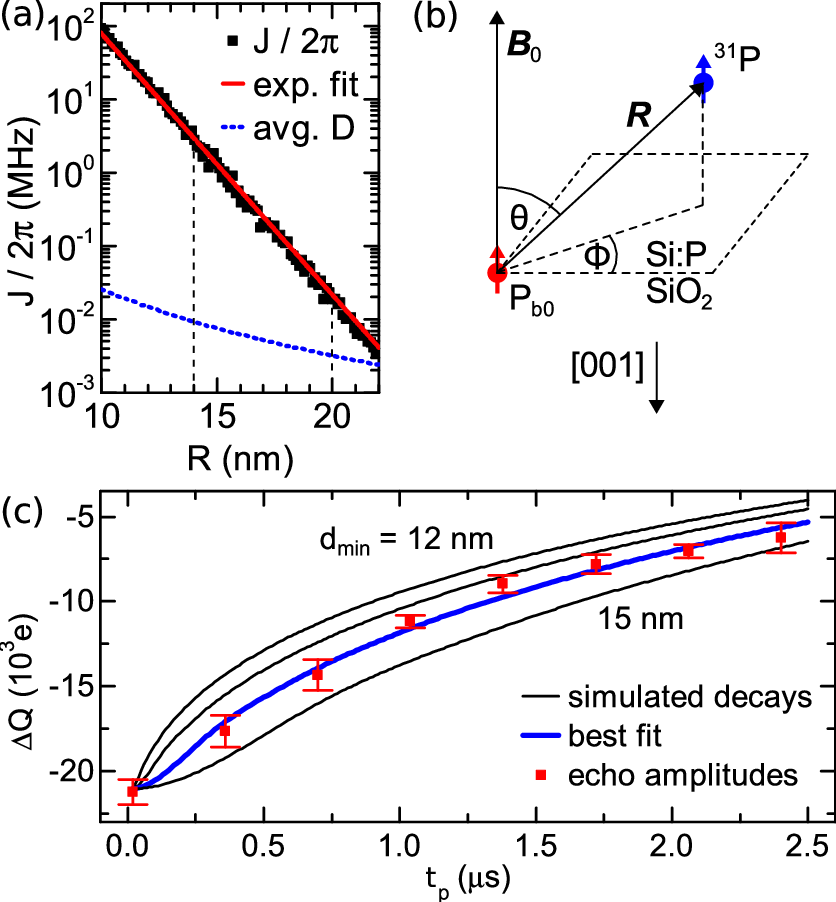}
\end{center}
\caption{\label{fig:simulationDecay}	
	(a)~Numerically calculated exchange interaction of $^{31}$P and P$_{\text{b0}}$ as a function of $R$. The result can be described by an exponential fit with decay constant $a_{\text{B}}^*/2$ (red solid line). For comparison the dipole-dipole coupling is shown as well (blue dashed line). The integration boundaries $d_{\text{min}}$ and $R_{\text{max}}$ for the best simulation of the experimental decay are marked by vertical dashed lines.
	(b)~Geometry used to describe the dipole-dipole coupling: The vector $\boldsymbol{R}$ connecting a dangling bond P$_{\text{b0}}$ and a donor $^{31}$P and the magnetic field $\boldsymbol{B}_0$ enclose the angle $\theta$. The orientation of the projection of $\boldsymbol{R}$ into the plane of the Si/SiO$_2$ interface is given by the angle $\phi$.
	(c)~Simulated EDDEER echo amplitude decays using $R_{\text{max}} = 20\,$nm and $d_{\text{min}} = 12$\,nm to 15\,nm (in steps of 1\,nm) compared to an experimentally recorded EDDEER echo amplitude decay (red squares). The best fit (blue line) is obtained for $d_{\text{min}} = 14$\,nm.
}
\end{figure}

We calculate the exchange interaction with a Heitler-London approach~\cite{Hei27,Sug27} as the energetic difference of the singlet and triplet states.
The ground-state wavefunction of the $^{31}$P electron located at $\boldsymbol{R}$ is modeled as an isotropic hydrogen-like orbital
\begin{equation}
	\psi (\boldsymbol{r}) = \frac{1}{\sqrt{\pi a^*_{\text{B}}}} \exp\left(-|\boldsymbol{r} - \boldsymbol{R}|/a^*_{\text{B}}\right)
\label{eq:wavefunctions}
\end{equation}
with the effective Bohr radius $a^*_{\text{B}}$.
The electron effective mass $m^* = 0.26\,m_0 = 3 (1/m_\parallel + 2/m_\perp)^{-1}$ (with $m_\parallel = 0.98\,m_0$ and $m_\perp = 0.19\,m_0$) and dielectric constant $\epsilon = 11.7$ in the silicon crystal yield an effective Bohr radius $a^*_{\text{B}} = 2.4$\,nm~\cite{Sze07}.
The Bloch character is not included in the simulations since its effect is averaged out for a random distribution of donors~\cite{Koi01}.
The P$_{\text{b0}}$ wavefunction is mainly localized at the respective Si atom as confirmed by measurements of the hyperfine interaction with the nearest neighbour nuclei~\cite{Ste98a}.
We therefore use a highly localized hydrogen-like orbital with an effective radius of half the Si-Si bond-length $a^*_{\text{db}}\approx 1.2$\,\r{A}\ as a simplified model of the P$_{\text{b0}}$ wave function~\cite{Hoe11}.
We calculate the exchange interaction as a function of $R$ solving the integrals numerically with Monte Carlo integration~\cite{Pre07}.
The result shown in Figure~\ref{fig:simulationDecay}\,(a) (squares) can be described by an exponential decay with decay constant $a^*_{\text{B}}/2$ (red line).
For comparison, we also plot the dipole-dipole coupling averaged over all spin pairs with a spin-spin distance $R$ with the $^{31}$P donors spin located on the surface of a hemisphere, described by the angle $\theta$ between the vector $\boldsymbol{R}$ connecting the two spins and the magnetic field and the azimuth angle $\phi$ [see Figure~\ref{fig:simulationDecay}\,(b)] as a function of $R$ [blue dashed line in Figure~\ref{fig:simulationDecay}~(a)], given by $D(R) \approx 26\,\text{MHz}\cdot\text{nm}^3/R^3$~\cite{Jes02}. 
The dipole-dipole coupling is much smaller than the exchange interaction for all distances $\leq 20$\,nm, which corresponds to the thickness of the doped epilayer studied, and we therefore neglect its contribution to the $^{31}$P-P$_{\text{b0}}$ coupling.

We further calculate the expected EDDEER response $\Delta Q (t_{\text{p}})$ by averaging the oscillations $\cos[J(R)t_{\text{p}}]$ over a distribution $\rho(R)$ of $^{31}$P-P$_{\text{b0}}$ distances 
\begin{equation}
	\Delta Q(t_{\text{p}}) = \Delta Q(0) \frac{\int \text{d}R {\rho(R) \cos[J(R) t_{\text{p}}]}}{\int \text{d}R \rho(R)}
\label{eq:sumOscillations}
\end{equation}
where $\Delta Q(0)$ is the pulsed EDMR echo amplitude for $t_{\text{p}} = 0$.
The integration area shown in grey in Figure~\ref{fig:integrationGeometry} is defined by two parameters, the distance $d_{\text{min}}$ from the Si/SiO$_2$ interface and the radius $R_{\text{max}}$ around the P$_{\text{b0}}$ center.
We assume a constant density of $^{31}$P donors within the 20 nm-thick epilayer for the average over all observed P$_{\text{b0}}$ centers, so that $\rho(R) \propto R^2$.
If $R_{\text{max}}$ exceeds the width of the doped layer, the integration area is cut accordingly.

The integration boundaries $d_{\text{min}}$ and  $R_{\text{max}}$ used in our model (Figure \ref{fig:integrationGeometry}) are a consequence of several constraints concerning the conditions for which pEDMR signals can be observed. 
Most importantly, the spin allowed transition rate 1/$\tau_{\text{ap}}$ of electrons between $^{31}$P donors and P$_{\text{b0}}$ centers~\cite{Dre12} is expected to depend on the distance $R$ between them~\cite{Tho65}.
For spin pairs which are too close, the recombination time constant becomes shorter than the free evolution interval of the spin echo and, therefore, these spin pairs do not contribute to the EDDEER signal.
For the EDDEER experiments shown above, $2\tau_1 = 5$\,$\mu$s, so that spin pairs with $\tau_{\text{ap}} \ll 5\,\mu$s  will not be observed. 
This lower bound of $\tau_{\text{ap}}$ corresponds to a minimum distance $R_{\text{min}}$ between the recombination partners. 
The typical density of dangling bonds at the Si/SiO$_2$ interface is 10$^{12}$\,cm$^{-2}$ for a native oxide~\cite{Pie02}, corresponding to an average distance of $\sim$10\,nm between the P$_{\text{b0}}$ centers.
If the average P$_{\text{b0}}$-P$_{\text{b0}}$ distance is smaller than $R_{\text{min}}$, the overlap of the capture volumes will lead to an effective layer of thickness $d_{\text{min}}$ in which all $^{31}$P recombine so quickly that they are not observed in our experiment.
As we will see below, this case indeed holds in our samples and we therefore use a minimum distance from the interface $d_{\text{min}}$ rather than a minimum distance $R_{\text{min}}$ from the considered $P_{\text{b0}}$ center as a lower boundary for the integration interval.
In contrast, for spin pairs with too large values of $R$, the recombination time constants become so long that no recombination occurs during the measurement time interval given by the upper bound of 40\,$\mu$s of the box-car integration interval, so that spin pairs with $\tau_{\text{ap}} \gg 40\,\mu$s also do not contribute to the EDDEER signal resulting in a maximum spin-spin distance $R_{\text{max}}$.
The recombination time constants which are observed in the EDDEER experiment presented, therefore, span a range of more than one order of magnitude.

The experimentally observed decay [red squares in Figure~\ref{fig:simulationDecay}\,(c)] is best described by a simulation using Equation~\eqref{eq:sumOscillations} for a distribution of $^{31}$P-P$_{\text{b0}}$ distances from $d_{\text{min}} = 14$\,nm to $R_{\text{max}} = 20$\,nm, as determined by a least-squares fit (blue line) with $d_{\text{min}}$ and $R_{\text{max}}$ as free parameters.
This corresponds to an average over individual values of the exchange coupling $25\,\text{kHz}\le{J}/2\pi\le 3$\,MHz.
For comparison, further simulations with $d_{\text{min}}$ ranging from 12\,nm to 15\,nm in steps of 1\,nm and $R_{\text{max}} = 20$\,nm are shown as well demonstrating that the resulting decay is rather sensitive to variations in $d_{\text{min}}$.

\section{Discussion and conclusions}
To further support our model, we estimate the dependence of $\tau_{\text{ap}}$ on $R$ by assuming that the recombination process involves an electron tunneling process through a  potential barrier between $^{31}$P and P$_{\text{b0}}$.
Using the WKB-method~\cite{Coh77} we estimate
\begin{equation}
	\frac{1}{\tau_{\text{ap}}} \propto \exp \left(-\frac{2}{\hbar}\int_0^R \text{d}x \sqrt{2m^*\Delta V(x)}\right),
\label{eq:tau_ap}
\end{equation}
where a barrier width $R$ corresponding to the $^{31}$P-P$_{\text{b0}}$ distance and flat bands have been assumed. A first estimate for the potential barrier could be $\Delta V(x) = 45$\,meV, corresponding to the binding energy of an electron in the $^{31}$P donor~\cite{Sze07}, so that Equation~\eqref{eq:tau_ap} becomes $1/\tau_{\text{ap}} \propto \exp[-R/(a^*_{\text{B}}/2)]$~\cite{Tho65}.
However, a comparison of the distances $d_{\text{min}}$ and $R_{\text{max}}$ and the variation of $\tau_{\text{ap}}$ estimated for the EDDEER pulse sequence length and the upper boxcar integration bounds indicates a significantly lower barrier of about 20\,meV if the recombination can indeed be described by a WKB-model. 
This significantly lower barrier suggests that a more realistic model of the tunneling process has to account for the Coulomb potential of the phosphorus donor and the binding energy of the dangling bond.
%
\begin{figure}[htb]
\begin{center}
\includegraphics{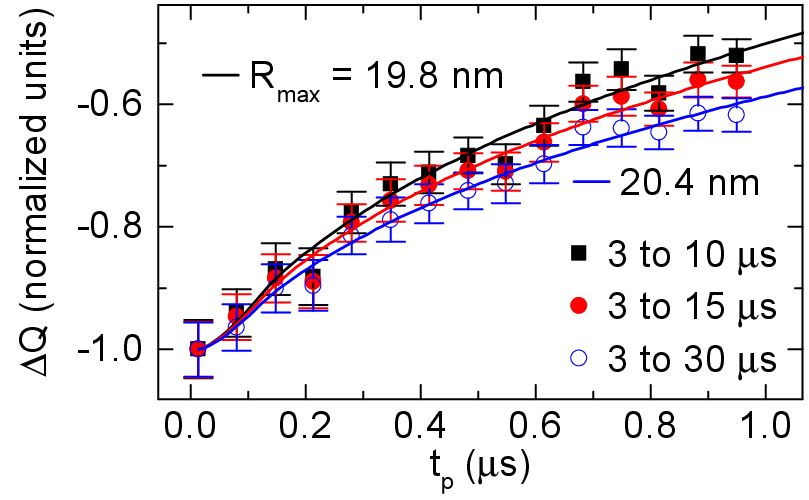}
\end{center}
\caption{\label{fig:effectIntegrationBound}	
	EDDEER echo amplitudes extracted from the experimentally recorded transients with boxcar integration intervals ranging from 3\,$\mu$s to 10\,$\mu$s (full squares), 3\,$\mu$s to 15\,$\mu$s (full circles) and 3\,$\mu$s to 30\,$\mu$s (open circles) with best fits obtained for $d_{\text{min}} = 13.2$\,nm and $R_{\text{max}} = 19.8$\,nm, 20\,nm and 20.4\,nm, respectively (solid lines). We here used $\tau_1 = 1$\,$\mu$s instead of $\tau_1 = 2.5\,\mu$s to obtain a better signal-to-noise ratio.
}
\end{figure}

Using Equation~\eqref{eq:tau_ap}, we can still obtain an estimate of the expected change of the upper bound $R_{\text{max}}$ of the observed range of $^{31}$P-P$_{\text{b0}}$ distances caused by a variation of the upper bound of the boxcar integration interval and compare it with experimental results.
To this end, we recorded additional EDDEER decays for boxcar integration intervals ranging from 3\,$\mu$s to 10\,$\mu$s (full squares in Figure~\ref{fig:effectIntegrationBound}), 3\,$\mu$s to 15\,$\mu$s (full circles) and 3\,$\mu$s to 30\,$\mu$s (open circles), respectively, and fitted the results as described above.
As expected, for an increase of the upper bound of the boxcar integration interval slower decays are observed corresponding to larger $R_{\text{max}}$.
We find that the respective EDDEER decays are best described by a simulation of the signals (solid lines) with $R_{\text{max}} = 19.8$\,nm, 20\,nm and 20.4\,nm in reasonable agreement with the variation of 1.4\,nm estimated using Equation~\eqref{eq:tau_ap} with $\Delta V(x) = 20$\,meV.

The total free evolution time of the spin echo $2\tau_1$ is longer than the $T_2$ time of the spins~\cite{Hue08}, so that only a small subensemble of $\sim$2\% of the spin pairs contributes to the EDDEER signal also taking into account above $d_{\text{min}}$.
However, we assume that the $T_2$ time of the P$_{\text{b0}}$ spins does not depend on the spin pair distance, so that for this subensemble the range of coupling constants is the same as for the whole spin pair ensemble.
Indeed, shortening $\tau_1$ to $1\,\mu$s as in Figure~\ref{fig:effectIntegrationBound} leads to essentially the same EDDEER decay time constant.

Based on EDMR experiments some estimates of the spin-spin distances have been reported in the literature.
The range of exchange coupling constants found here corresponds well to previous estimates for spin pairs formed by $^{31}$P donors and radiation defects (SL1), where an upper bound of the $^{31}$P-SL1 distance of 20\,nm has been reported~\cite{Akh11}.
The range of coupling constants also compares favorably to the upper bound of $J/2\pi < 5$\,MHz for $^{31}$P-P$_{\text{b0}}$ spin pairs obtained by Lu et al.~\cite{Lu11}. 
In contrast, a much smaller spin-spin distance of $R \approx 4\,$nm has been reported based on the observed $^{31}$P decoherence time induced by magnetic field fluctuation at the Si/SiO$_2$ interface~\cite{Pai10}.
However, these authors have evaluated their experimental data with a model by de Sousa~\cite{Sou07} and mention that not all requirements for the application of the model were met.
Using our estimation of the exchange coupling, $R \approx 4\,$nm would correspond to $J/2\pi \approx 13 $\,GHz which is not compatible with the experimental data.

The results obtained here suggest a more systematic study of the EDDEER decay for samples with different thicknesses of the doped epilayer and, therefore, different distributions of exchange coupling requiring larger changes of the boxcar integration interval than presented in the proof-of-principle experiments presented here.
Ultimately, a $\delta$-doped layer of $^{31}$P donors~\cite{Mc09} could provide a much better defined coupling between the donors and the P$_{\text{b0}}$ defects.

To summarize, we have measured the exchange coupling between $^{31}$P donors and P$_{\text{b0}}$ defects at the Si/SiO$_2$ interface using electrically detected DEER.
We find that the measured EDDEER signal can be explained by a distribution of $^{31}$P-P$_{\text{b0}}$ coupling strengths resulting from a distribution of $^{31}$P-P$_{\text{b0}}$ distances over the observed spin ensemble.
A simulation of the experimental signal with a numerical calculation of the exchange coupling allows us to understand the experimental results by a distribution of $^{31}$P-P$_{\text{b0}}$ distances ranging from 14\,nm to 20\,nm corresponding to an exchange coupling from $25$\,kHz to $3$\,MHz.

\section*{Acknowledgements}
The work was supported by DFG (Grant No. SFB 631, C3 and Grant No. SPP 1601, Br 1585/8-1) and BMBF (Grant No. EPR Solar).

\bibliographystyle{apsrev4-1.bst}
\bibliography{literature}

\begin{thebibliography}{40}%
\makeatletter
\providecommand \@ifxundefined [1]{%
 \@ifx{#1\undefined}
}%
\providecommand \@ifnum [1]{%
 \ifnum #1\expandafter \@firstoftwo
 \else \expandafter \@secondoftwo
 \fi
}%
\providecommand \@ifx [1]{%
 \ifx #1\expandafter \@firstoftwo
 \else \expandafter \@secondoftwo
 \fi
}%
\providecommand \natexlab [1]{#1}%
\providecommand \enquote  [1]{``#1''}%
\providecommand \bibnamefont  [1]{#1}%
\providecommand \bibfnamefont [1]{#1}%
\providecommand \citenamefont [1]{#1}%
\providecommand \href@noop [0]{\@secondoftwo}%
\providecommand \href [0]{\begingroup \@sanitize@url \@href}%
\providecommand \@href[1]{\@@startlink{#1}\@@href}%
\providecommand \@@href[1]{\endgroup#1\@@endlink}%
\providecommand \@sanitize@url [0]{\catcode `\\12\catcode `\$12\catcode
  `\&12\catcode `\#12\catcode `\^12\catcode `\_12\catcode `\%12\relax}%
\providecommand \@@startlink[1]{}%
\providecommand \@@endlink[0]{}%
\providecommand \url  [0]{\begingroup\@sanitize@url \@url }%
\providecommand \@url [1]{\endgroup\@href {#1}{\urlprefix }}%
\providecommand \urlprefix  [0]{URL }%
\providecommand \Eprint [0]{\href }%
\providecommand \doibase [0]{http://dx.doi.org/}%
\providecommand \selectlanguage [0]{\@gobble}%
\providecommand \bibinfo  [0]{\@secondoftwo}%
\providecommand \bibfield  [0]{\@secondoftwo}%
\providecommand \translation [1]{[#1]}%
\providecommand \BibitemOpen [0]{}%
\providecommand \bibitemStop [0]{}%
\providecommand \bibitemNoStop [0]{.\EOS\space}%
\providecommand \EOS [0]{\spacefactor3000\relax}%
\providecommand \BibitemShut  [1]{\csname bibitem#1\endcsname}%
\let\auto@bib@innerbib\@empty
\bibitem [{\citenamefont {Spaeth}\ and\ \citenamefont {Overhof}(2003)}]{Spa03}%
  \BibitemOpen
  \bibfield  {author} {\bibinfo {author} {\bibfnamefont {J.-M.}\ \bibnamefont
  {Spaeth}}\ and\ \bibinfo {author} {\bibfnamefont {H.}~\bibnamefont
  {Overhof}},\ }\href@noop {} {\emph {\bibinfo {title} {Point Defects in
  Semiconductors and Insulators}}}\ (\bibinfo  {publisher} {Springer},\
  \bibinfo {address} {Berlin},\ \bibinfo {year} {2003})\BibitemShut {NoStop}%
\bibitem [{\citenamefont {Lepine}(1972)}]{Lep72}%
  \BibitemOpen
  \bibfield  {author} {\bibinfo {author} {\bibfnamefont {D.~J.}\ \bibnamefont
  {Lepine}},\ }\href@noop {} {\bibfield  {journal} {\bibinfo  {journal} {Phys.
  Rev. B}\ }\textbf {\bibinfo {volume} {6}},\ \bibinfo {pages} {436} (\bibinfo
  {year} {1972})}\BibitemShut {NoStop}%
\bibitem [{\citenamefont {Brandt}\ \emph {et~al.}(2004)\citenamefont {Brandt},
  \citenamefont {Goennenwein}, \citenamefont {Graf}, \citenamefont {Huebl},
  \citenamefont {Lauterbach},\ and\ \citenamefont {Stutzmann}}]{Bra04}%
  \BibitemOpen
  \bibfield  {author} {\bibinfo {author} {\bibfnamefont {M.~S.}\ \bibnamefont
  {Brandt}}, \bibinfo {author} {\bibfnamefont {S.~T.~B.}\ \bibnamefont
  {Goennenwein}}, \bibinfo {author} {\bibfnamefont {T.}~\bibnamefont {Graf}},
  \bibinfo {author} {\bibfnamefont {H.}~\bibnamefont {Huebl}}, \bibinfo
  {author} {\bibfnamefont {S.}~\bibnamefont {Lauterbach}}, \ and\ \bibinfo
  {author} {\bibfnamefont {M.}~\bibnamefont {Stutzmann}},\ }\href@noop {}
  {\bibfield  {journal} {\bibinfo  {journal} {phys. stat. sol. (c)}\ }\textbf
  {\bibinfo {volume} {1}},\ \bibinfo {pages} {2056} (\bibinfo {year}
  {2004})}\BibitemShut {NoStop}%
\bibitem [{\citenamefont {{McCamey}}\ \emph {et~al.}(2006)\citenamefont
  {{McCamey}}, \citenamefont {Huebl}, \citenamefont {Brandt}, \citenamefont
  {Hutchison}, \citenamefont {{McCallum}}, \citenamefont {Clark},\ and\
  \citenamefont {Hamilton}}]{Mc06}%
  \BibitemOpen
  \bibfield  {author} {\bibinfo {author} {\bibfnamefont {D.~R.}\ \bibnamefont
  {{McCamey}}}, \bibinfo {author} {\bibfnamefont {H.}~\bibnamefont {Huebl}},
  \bibinfo {author} {\bibfnamefont {M.~S.}\ \bibnamefont {Brandt}}, \bibinfo
  {author} {\bibfnamefont {W.~D.}\ \bibnamefont {Hutchison}}, \bibinfo {author}
  {\bibfnamefont {J.~C.}\ \bibnamefont {{McCallum}}}, \bibinfo {author}
  {\bibfnamefont {R.~G.}\ \bibnamefont {Clark}}, \ and\ \bibinfo {author}
  {\bibfnamefont {A.~R.}\ \bibnamefont {Hamilton}},\ }\href@noop {} {\bibfield
  {journal} {\bibinfo  {journal} {Appl. Phys. Lett.}\ }\textbf {\bibinfo
  {volume} {89}},\ \bibinfo {pages} {182115} (\bibinfo {year}
  {2006})}\BibitemShut {NoStop}%
\bibitem [{\citenamefont {Baker}\ \emph {et~al.}(2012)\citenamefont {Baker},
  \citenamefont {Ambal}, \citenamefont {Waters}, \citenamefont {Baarda},
  \citenamefont {Morishita}, \citenamefont {van Schooten}, \citenamefont
  {McCamey}, \citenamefont {Lupton},\ and\ \citenamefont {Boehme}}]{Bak12}%
  \BibitemOpen
  \bibfield  {author} {\bibinfo {author} {\bibfnamefont {W.}~\bibnamefont
  {Baker}}, \bibinfo {author} {\bibfnamefont {K.}~\bibnamefont {Ambal}},
  \bibinfo {author} {\bibfnamefont {D.}~\bibnamefont {Waters}}, \bibinfo
  {author} {\bibfnamefont {R.}~\bibnamefont {Baarda}}, \bibinfo {author}
  {\bibfnamefont {H.}~\bibnamefont {Morishita}}, \bibinfo {author}
  {\bibfnamefont {K.}~\bibnamefont {van Schooten}}, \bibinfo {author}
  {\bibfnamefont {D.}~\bibnamefont {McCamey}}, \bibinfo {author} {\bibfnamefont
  {J.}~\bibnamefont {Lupton}}, \ and\ \bibinfo {author} {\bibfnamefont
  {C.}~\bibnamefont {Boehme}},\ }\href@noop {} {\bibfield  {journal} {\bibinfo
  {journal} {Nature Commun.}\ }\textbf {\bibinfo {volume} {3}},\ \bibinfo
  {pages} {898} (\bibinfo {year} {2012})}\BibitemShut {NoStop}%
\bibitem [{\citenamefont {Schnegg}\ \emph {et~al.}(2012)\citenamefont
  {Schnegg}, \citenamefont {Behrends}, \citenamefont {Fehr},\ and\
  \citenamefont {Lips}}]{Sch12}%
  \BibitemOpen
  \bibfield  {author} {\bibinfo {author} {\bibfnamefont {A.}~\bibnamefont
  {Schnegg}}, \bibinfo {author} {\bibfnamefont {J.}~\bibnamefont {Behrends}},
  \bibinfo {author} {\bibfnamefont {M.}~\bibnamefont {Fehr}}, \ and\ \bibinfo
  {author} {\bibfnamefont {K.}~\bibnamefont {Lips}},\ }\href@noop {} {\bibfield
   {journal} {\bibinfo  {journal} {Phys. Chem. Chem. Phys.}\ }\textbf {\bibinfo
  {volume} {14}},\ \bibinfo {pages} {14418} (\bibinfo {year}
  {2012})}\BibitemShut {NoStop}%
\bibitem [{\citenamefont {Schmidt}\ and\ \citenamefont
  {Solomon}(1966)}]{Sch66}%
  \BibitemOpen
  \bibfield  {author} {\bibinfo {author} {\bibfnamefont {J.}~\bibnamefont
  {Schmidt}}\ and\ \bibinfo {author} {\bibfnamefont {I.}~\bibnamefont
  {Solomon}},\ }\href@noop {} {\bibfield  {journal} {\bibinfo  {journal}
  {Compt. Rend.}\ }\textbf {\bibinfo {volume} {B 263}},\ \bibinfo {pages} {169}
  (\bibinfo {year} {1966})}\BibitemShut {NoStop}%
\bibitem [{\citenamefont {Kaplan}\ \emph {et~al.}(1978)\citenamefont {Kaplan},
  \citenamefont {Solomon},\ and\ \citenamefont {Mott}}]{Kap78}%
  \BibitemOpen
  \bibfield  {author} {\bibinfo {author} {\bibfnamefont {D.}~\bibnamefont
  {Kaplan}}, \bibinfo {author} {\bibfnamefont {I.}~\bibnamefont {Solomon}}, \
  and\ \bibinfo {author} {\bibfnamefont {N.}~\bibnamefont {Mott}},\ }\href@noop
  {} {\bibfield  {journal} {\bibinfo  {journal} {J. Phys. Lett. (Paris)}\
  }\textbf {\bibinfo {volume} {39}},\ \bibinfo {pages} {51} (\bibinfo {year}
  {1978})}\BibitemShut {NoStop}%
\bibitem [{\citenamefont {Kishimoto}\ \emph {et~al.}(1981)\citenamefont
  {Kishimoto}, \citenamefont {Morigaki},\ and\ \citenamefont
  {Murakami}}]{Kis81}%
  \BibitemOpen
  \bibfield  {author} {\bibinfo {author} {\bibfnamefont {N.}~\bibnamefont
  {Kishimoto}}, \bibinfo {author} {\bibfnamefont {K.}~\bibnamefont {Morigaki}},
  \ and\ \bibinfo {author} {\bibfnamefont {K.}~\bibnamefont {Murakami}},\
  }\href@noop {} {\bibfield  {journal} {\bibinfo  {journal} {J. Phys. Soc.
  Jpn.}\ }\textbf {\bibinfo {volume} {50}},\ \bibinfo {pages} {1970} (\bibinfo
  {year} {1981})}\BibitemShut {NoStop}%
\bibitem [{\citenamefont {Stich}\ \emph {et~al.}(1995)\citenamefont {Stich},
  \citenamefont {{Greulich-Weber}},\ and\ \citenamefont {Spaeth}}]{Sti95}%
  \BibitemOpen
  \bibfield  {author} {\bibinfo {author} {\bibfnamefont {B.}~\bibnamefont
  {Stich}}, \bibinfo {author} {\bibfnamefont {S.}~\bibnamefont
  {{Greulich-Weber}}}, \ and\ \bibinfo {author} {\bibfnamefont {J.-M.}\
  \bibnamefont {Spaeth}},\ }\href@noop {} {\bibfield  {journal} {\bibinfo
  {journal} {J. Appl. Phys.}\ }\textbf {\bibinfo {volume} {77}},\ \bibinfo
  {pages} {1546} (\bibinfo {year} {1995})}\BibitemShut {NoStop}%
\bibitem [{\citenamefont {Hoehne}\ \emph {et~al.}(2010)\citenamefont {Hoehne},
  \citenamefont {Huebl}, \citenamefont {Galler}, \citenamefont {Stutzmann},\
  and\ \citenamefont {Brandt}}]{Hoe10}%
  \BibitemOpen
  \bibfield  {author} {\bibinfo {author} {\bibfnamefont {F.}~\bibnamefont
  {Hoehne}}, \bibinfo {author} {\bibfnamefont {H.}~\bibnamefont {Huebl}},
  \bibinfo {author} {\bibfnamefont {B.}~\bibnamefont {Galler}}, \bibinfo
  {author} {\bibfnamefont {M.}~\bibnamefont {Stutzmann}}, \ and\ \bibinfo
  {author} {\bibfnamefont {M.~S.}\ \bibnamefont {Brandt}},\ }\href@noop {}
  {\bibfield  {journal} {\bibinfo  {journal} {Phys. Rev. Lett.}\ }\textbf
  {\bibinfo {volume} {104}},\ \bibinfo {pages} {046402} (\bibinfo {year}
  {2010})}\BibitemShut {NoStop}%
\bibitem [{\citenamefont {Kane}(1998)}]{Kan98}%
  \BibitemOpen
  \bibfield  {author} {\bibinfo {author} {\bibfnamefont {B.~E.}\ \bibnamefont
  {Kane}},\ }\href@noop {} {\bibfield  {journal} {\bibinfo  {journal} {Nature}\
  }\textbf {\bibinfo {volume} {393}},\ \bibinfo {pages} {133} (\bibinfo {year}
  {1998})}\BibitemShut {NoStop}%
\bibitem [{\citenamefont {Morello}\ \emph {et~al.}(2010)\citenamefont
  {Morello}, \citenamefont {Pla}, \citenamefont {Zwanenburg}, \citenamefont
  {Chan}, \citenamefont {Tan}, \citenamefont {Huebl}, \citenamefont {Mottonen},
  \citenamefont {Nugroho}, \citenamefont {Yang}, \citenamefont {van Donkelaar},
  \citenamefont {Alves}, \citenamefont {Jamieson}, \citenamefont {Escott},
  \citenamefont {Hollenberg}, \citenamefont {Clark},\ and\ \citenamefont
  {Dzurak}}]{Mor10}%
  \BibitemOpen
  \bibfield  {author} {\bibinfo {author} {\bibfnamefont {A.}~\bibnamefont
  {Morello}}, \bibinfo {author} {\bibfnamefont {J.~J.}\ \bibnamefont {Pla}},
  \bibinfo {author} {\bibfnamefont {F.~A.}\ \bibnamefont {Zwanenburg}},
  \bibinfo {author} {\bibfnamefont {K.~W.}\ \bibnamefont {Chan}}, \bibinfo
  {author} {\bibfnamefont {K.~Y.}\ \bibnamefont {Tan}}, \bibinfo {author}
  {\bibfnamefont {H.}~\bibnamefont {Huebl}}, \bibinfo {author} {\bibfnamefont
  {M.}~\bibnamefont {Mottonen}}, \bibinfo {author} {\bibfnamefont {C.~D.}\
  \bibnamefont {Nugroho}}, \bibinfo {author} {\bibfnamefont {C.}~\bibnamefont
  {Yang}}, \bibinfo {author} {\bibfnamefont {J.~A.}\ \bibnamefont {van
  Donkelaar}}, \bibinfo {author} {\bibfnamefont {A.~D.~C.}\ \bibnamefont
  {Alves}}, \bibinfo {author} {\bibfnamefont {D.~N.}\ \bibnamefont {Jamieson}},
  \bibinfo {author} {\bibfnamefont {C.~C.}\ \bibnamefont {Escott}}, \bibinfo
  {author} {\bibfnamefont {L.~C.~L.}\ \bibnamefont {Hollenberg}}, \bibinfo
  {author} {\bibfnamefont {R.~G.}\ \bibnamefont {Clark}}, \ and\ \bibinfo
  {author} {\bibfnamefont {A.~S.}\ \bibnamefont {Dzurak}},\ }\href@noop {}
  {\bibfield  {journal} {\bibinfo  {journal} {Nature}\ }\textbf {\bibinfo
  {volume} {467}},\ \bibinfo {pages} {687} (\bibinfo {year}
  {2010})}\BibitemShut {NoStop}%
\bibitem [{\citenamefont {Boehme}\ and\ \citenamefont {Lips}(2002)}]{Boe02}%
  \BibitemOpen
  \bibfield  {author} {\bibinfo {author} {\bibfnamefont {C.}~\bibnamefont
  {Boehme}}\ and\ \bibinfo {author} {\bibfnamefont {K.}~\bibnamefont {Lips}},\
  }\href@noop {} {\bibfield  {journal} {\bibinfo  {journal} {phys. stat. sol.
  (b)}\ }\textbf {\bibinfo {volume} {233}},\ \bibinfo {pages} {427} (\bibinfo
  {year} {2002})}\BibitemShut {NoStop}%
\bibitem [{\citenamefont {de~Sousa}(2007)}]{Sou07}%
  \BibitemOpen
  \bibfield  {author} {\bibinfo {author} {\bibfnamefont {R.}~\bibnamefont
  {de~Sousa}},\ }\href@noop {} {\bibfield  {journal} {\bibinfo  {journal}
  {Phys. Rev. B}\ }\textbf {\bibinfo {volume} {76}},\ \bibinfo {pages} {245306}
  (\bibinfo {year} {2007})}\BibitemShut {NoStop}%
\bibitem [{\citenamefont {Stegner}\ \emph {et~al.}(2006)\citenamefont
  {Stegner}, \citenamefont {Boehme}, \citenamefont {Huebl}, \citenamefont
  {Stutzmann}, \citenamefont {Lips},\ and\ \citenamefont {Brandt}}]{Ste06}%
  \BibitemOpen
  \bibfield  {author} {\bibinfo {author} {\bibfnamefont {A.~R.}\ \bibnamefont
  {Stegner}}, \bibinfo {author} {\bibfnamefont {C.}~\bibnamefont {Boehme}},
  \bibinfo {author} {\bibfnamefont {H.}~\bibnamefont {Huebl}}, \bibinfo
  {author} {\bibfnamefont {M.}~\bibnamefont {Stutzmann}}, \bibinfo {author}
  {\bibfnamefont {K.}~\bibnamefont {Lips}}, \ and\ \bibinfo {author}
  {\bibfnamefont {M.~S.}\ \bibnamefont {Brandt}},\ }\href@noop {} {\bibfield
  {journal} {\bibinfo  {journal} {Nature Phys.}\ }\textbf {\bibinfo {volume}
  {2}},\ \bibinfo {pages} {835} (\bibinfo {year} {2006})}\BibitemShut {NoStop}%
\bibitem [{\citenamefont {Huebl}\ \emph {et~al.}(2008)\citenamefont {Huebl},
  \citenamefont {Hoehne}, \citenamefont {Grolik}, \citenamefont {Stegner},
  \citenamefont {Stutzmann},\ and\ \citenamefont {Brandt}}]{Hue08}%
  \BibitemOpen
  \bibfield  {author} {\bibinfo {author} {\bibfnamefont {H.}~\bibnamefont
  {Huebl}}, \bibinfo {author} {\bibfnamefont {F.}~\bibnamefont {Hoehne}},
  \bibinfo {author} {\bibfnamefont {B.}~\bibnamefont {Grolik}}, \bibinfo
  {author} {\bibfnamefont {A.~R.}\ \bibnamefont {Stegner}}, \bibinfo {author}
  {\bibfnamefont {M.}~\bibnamefont {Stutzmann}}, \ and\ \bibinfo {author}
  {\bibfnamefont {M.~S.}\ \bibnamefont {Brandt}},\ }\href@noop {} {\bibfield
  {journal} {\bibinfo  {journal} {Phys. Rev. Lett.}\ }\textbf {\bibinfo
  {volume} {100}},\ \bibinfo {pages} {177602} (\bibinfo {year}
  {2008})}\BibitemShut {NoStop}%
\bibitem [{\citenamefont {Dreher}\ \emph {et~al.}(2012)\citenamefont {Dreher},
  \citenamefont {Hoehne}, \citenamefont {Stutzmann},\ and\ \citenamefont
  {Brandt}}]{Dre12}%
  \BibitemOpen
  \bibfield  {author} {\bibinfo {author} {\bibfnamefont {L.}~\bibnamefont
  {Dreher}}, \bibinfo {author} {\bibfnamefont {F.}~\bibnamefont {Hoehne}},
  \bibinfo {author} {\bibfnamefont {M.}~\bibnamefont {Stutzmann}}, \ and\
  \bibinfo {author} {\bibfnamefont {M.~S.}\ \bibnamefont {Brandt}},\
  }\href@noop {} {\bibfield  {journal} {\bibinfo  {journal} {Phys. Rev. Lett.}\
  }\textbf {\bibinfo {volume} {108}},\ \bibinfo {pages} {027602} (\bibinfo
  {year} {2012})}\BibitemShut {NoStop}%
\bibitem [{\citenamefont {Lu}\ \emph {et~al.}(2011)\citenamefont {Lu},
  \citenamefont {Hoehne}, \citenamefont {Stegner}, \citenamefont {Dreher},
  \citenamefont {Stutzmann}, \citenamefont {Brandt},\ and\ \citenamefont
  {Huebl}}]{Lu11}%
  \BibitemOpen
  \bibfield  {author} {\bibinfo {author} {\bibfnamefont {J.}~\bibnamefont
  {Lu}}, \bibinfo {author} {\bibfnamefont {F.}~\bibnamefont {Hoehne}}, \bibinfo
  {author} {\bibfnamefont {A.~R.}\ \bibnamefont {Stegner}}, \bibinfo {author}
  {\bibfnamefont {L.}~\bibnamefont {Dreher}}, \bibinfo {author} {\bibfnamefont
  {M.}~\bibnamefont {Stutzmann}}, \bibinfo {author} {\bibfnamefont {M.~S.}\
  \bibnamefont {Brandt}}, \ and\ \bibinfo {author} {\bibfnamefont
  {H.}~\bibnamefont {Huebl}},\ }\href@noop {} {\bibfield  {journal} {\bibinfo
  {journal} {Phys. Rev. B}\ }\textbf {\bibinfo {volume} {83}},\ \bibinfo
  {pages} {235201} (\bibinfo {year} {2011})}\BibitemShut {NoStop}%
\bibitem [{\citenamefont {Milov}\ \emph {et~al.}(1981)\citenamefont {Milov},
  \citenamefont {Salikhov},\ and\ \citenamefont {Shirov}}]{Mil81}%
  \BibitemOpen
  \bibfield  {author} {\bibinfo {author} {\bibfnamefont {A.~D.}\ \bibnamefont
  {Milov}}, \bibinfo {author} {\bibfnamefont {K.~M.}\ \bibnamefont {Salikhov}},
  \ and\ \bibinfo {author} {\bibfnamefont {M.~D.}\ \bibnamefont {Shirov}},\
  }\href@noop {} {\bibfield  {journal} {\bibinfo  {journal} {Sov. Phys. - Solid
  State}\ }\textbf {\bibinfo {volume} {24}},\ \bibinfo {pages} {565} (\bibinfo
  {year} {1981})}\BibitemShut {NoStop}%
\bibitem [{\citenamefont {Jeschke}(2002)}]{Jes02}%
  \BibitemOpen
  \bibfield  {author} {\bibinfo {author} {\bibfnamefont {G.}~\bibnamefont
  {Jeschke}},\ }\href@noop {} {\bibfield  {journal} {\bibinfo  {journal}
  {Macromol. Rapid Commun.}\ }\textbf {\bibinfo {volume} {23}},\ \bibinfo
  {pages} {227} (\bibinfo {year} {2002})}\BibitemShut {NoStop}%
\bibitem [{\citenamefont {Jeschke}\ and\ \citenamefont
  {Schweiger}(2001)}]{Jes01}%
  \BibitemOpen
  \bibfield  {author} {\bibinfo {author} {\bibfnamefont {G.}~\bibnamefont
  {Jeschke}}\ and\ \bibinfo {author} {\bibfnamefont {A.}~\bibnamefont
  {Schweiger}},\ }\href@noop {} {\emph {\bibinfo {title} {Principles of Pulse
  Electron Paramagnetic Resonance}}}\ (\bibinfo  {publisher} {Oxford University
  Press},\ \bibinfo {address} {Oxford},\ \bibinfo {year} {2001})\BibitemShut
  {NoStop}%
\bibitem [{\citenamefont {Klauder}\ and\ \citenamefont
  {Anderson}(1962)}]{Kla62}%
  \BibitemOpen
  \bibfield  {author} {\bibinfo {author} {\bibfnamefont {J.~R.}\ \bibnamefont
  {Klauder}}\ and\ \bibinfo {author} {\bibfnamefont {P.~W.}\ \bibnamefont
  {Anderson}},\ }\href@noop {} {\bibfield  {journal} {\bibinfo  {journal}
  {Phys. Rev.}\ }\textbf {\bibinfo {volume} {125}},\ \bibinfo {pages} {912}
  (\bibinfo {year} {1962})}\BibitemShut {NoStop}%
\bibitem [{\citenamefont {Boehme}\ and\ \citenamefont {Lips}(2003)}]{Boe03}%
  \BibitemOpen
  \bibfield  {author} {\bibinfo {author} {\bibfnamefont {C.}~\bibnamefont
  {Boehme}}\ and\ \bibinfo {author} {\bibfnamefont {K.}~\bibnamefont {Lips}},\
  }\href@noop {} {\bibfield  {journal} {\bibinfo  {journal} {Phys. Rev. B}\
  }\textbf {\bibinfo {volume} {68}},\ \bibinfo {pages} {245105} (\bibinfo
  {year} {2003})}\BibitemShut {NoStop}%
\bibitem [{\citenamefont {Hoehne}\ \emph {et~al.}(2012)\citenamefont {Hoehne},
  \citenamefont {Dreher}, \citenamefont {Behrends}, \citenamefont {Fehr},
  \citenamefont {Huebl}, \citenamefont {Lips}, \citenamefont {Schnegg},
  \citenamefont {Suckert}, \citenamefont {Stutzmann},\ and\ \citenamefont
  {Brandt}}]{Hoe12}%
  \BibitemOpen
  \bibfield  {author} {\bibinfo {author} {\bibfnamefont {F.}~\bibnamefont
  {Hoehne}}, \bibinfo {author} {\bibfnamefont {L.}~\bibnamefont {Dreher}},
  \bibinfo {author} {\bibfnamefont {J.}~\bibnamefont {Behrends}}, \bibinfo
  {author} {\bibfnamefont {M.}~\bibnamefont {Fehr}}, \bibinfo {author}
  {\bibfnamefont {H.}~\bibnamefont {Huebl}}, \bibinfo {author} {\bibfnamefont
  {K.}~\bibnamefont {Lips}}, \bibinfo {author} {\bibfnamefont {A.}~\bibnamefont
  {Schnegg}}, \bibinfo {author} {\bibfnamefont {M.}~\bibnamefont {Suckert}},
  \bibinfo {author} {\bibfnamefont {M.}~\bibnamefont {Stutzmann}}, \ and\
  \bibinfo {author} {\bibfnamefont {M.~S.}\ \bibnamefont {Brandt}},\
  }\href@noop {} {\bibfield  {journal} {\bibinfo  {journal} {Rev. Sci.
  Instrum.}\ }\textbf {\bibinfo {volume} {83}},\ \bibinfo {pages} {043907}
  (\bibinfo {year} {2012})}\BibitemShut {NoStop}%
\bibitem [{\citenamefont {Stesmans}\ and\ \citenamefont
  {Afanas'ev}(1998)}]{Ste98}%
  \BibitemOpen
  \bibfield  {author} {\bibinfo {author} {\bibfnamefont {A.}~\bibnamefont
  {Stesmans}}\ and\ \bibinfo {author} {\bibfnamefont {V.~V.}\ \bibnamefont
  {Afanas'ev}},\ }\href@noop {} {\bibfield  {journal} {\bibinfo  {journal} {J.
  Appl. Phys.}\ }\textbf {\bibinfo {volume} {83}},\ \bibinfo {pages} {2449}
  (\bibinfo {year} {1998})}\BibitemShut {NoStop}%
\bibitem [{\citenamefont {Feher}(1959)}]{Feh59}%
  \BibitemOpen
  \bibfield  {author} {\bibinfo {author} {\bibfnamefont {G.}~\bibnamefont
  {Feher}},\ }\href@noop {} {\bibfield  {journal} {\bibinfo  {journal} {Phys.
  Rev.}\ }\textbf {\bibinfo {volume} {114}},\ \bibinfo {pages} {1219} (\bibinfo
  {year} {1959})}\BibitemShut {NoStop}%
\bibitem [{\citenamefont {Heitler}\ and\ \citenamefont {London}(1927)}]{Hei27}%
  \BibitemOpen
  \bibfield  {author} {\bibinfo {author} {\bibfnamefont {W.}~\bibnamefont
  {Heitler}}\ and\ \bibinfo {author} {\bibfnamefont {F.}~\bibnamefont
  {London}},\ }\href@noop {} {\bibfield  {journal} {\bibinfo  {journal} {Z.
  Phys. A}\ }\textbf {\bibinfo {volume} {44}},\ \bibinfo {pages} {455}
  (\bibinfo {year} {1927})}\BibitemShut {NoStop}%
\bibitem [{\citenamefont {Sugiura}(1927)}]{Sug27}%
  \BibitemOpen
  \bibfield  {author} {\bibinfo {author} {\bibfnamefont {Y.}~\bibnamefont
  {Sugiura}},\ }\href@noop {} {\bibfield  {journal} {\bibinfo  {journal} {Z.
  Phys. A}\ }\textbf {\bibinfo {volume} {45}},\ \bibinfo {pages} {484}
  (\bibinfo {year} {1927})}\BibitemShut {NoStop}%
\bibitem [{\citenamefont {Sze}\ and\ \citenamefont {Ng}(2007)}]{Sze07}%
  \BibitemOpen
  \bibfield  {author} {\bibinfo {author} {\bibfnamefont {S.~M.}\ \bibnamefont
  {Sze}}\ and\ \bibinfo {author} {\bibfnamefont {K.~K.}\ \bibnamefont {Ng}},\
  }\href@noop {} {\emph {\bibinfo {title} {Semiconductor Devices: Physics and
  Technology}}}\ (\bibinfo  {publisher} {Wiley--Interscience},\ \bibinfo
  {address} {Hoboken, NJ},\ \bibinfo {year} {2007})\BibitemShut {NoStop}%
\bibitem [{\citenamefont {Koiller}\ \emph {et~al.}(2001)\citenamefont
  {Koiller}, \citenamefont {Hu},\ and\ \citenamefont {Das~Sarma}}]{Koi01}%
  \BibitemOpen
  \bibfield  {author} {\bibinfo {author} {\bibfnamefont {B.}~\bibnamefont
  {Koiller}}, \bibinfo {author} {\bibfnamefont {X.}~\bibnamefont {Hu}}, \ and\
  \bibinfo {author} {\bibfnamefont {S.}~\bibnamefont {Das~Sarma}},\ }\href@noop
  {} {\bibfield  {journal} {\bibinfo  {journal} {Phys. Rev. Lett.}\ }\textbf
  {\bibinfo {volume} {88}},\ \bibinfo {pages} {027903} (\bibinfo {year}
  {2001})}\BibitemShut {NoStop}%
\bibitem [{\citenamefont {Stesmans}\ \emph {et~al.}(1998)\citenamefont
  {Stesmans}, \citenamefont {Nouwen},\ and\ \citenamefont
  {Afanas'ev}}]{Ste98a}%
  \BibitemOpen
  \bibfield  {author} {\bibinfo {author} {\bibfnamefont {A.}~\bibnamefont
  {Stesmans}}, \bibinfo {author} {\bibfnamefont {B.}~\bibnamefont {Nouwen}}, \
  and\ \bibinfo {author} {\bibfnamefont {V.~V.}\ \bibnamefont {Afanas'ev}},\
  }\href@noop {} {\bibfield  {journal} {\bibinfo  {journal} {Phys. Rev. B}\
  }\textbf {\bibinfo {volume} {58}},\ \bibinfo {pages} {15801} (\bibinfo {year}
  {1998})}\BibitemShut {NoStop}%
\bibitem [{\citenamefont {Hoehne}\ \emph {et~al.}(2011)\citenamefont {Hoehne},
  \citenamefont {Lu}, \citenamefont {Stegner}, \citenamefont {Stutzmann},
  \citenamefont {Brandt}, \citenamefont {Rohrm\"uller}, \citenamefont
  {Schmidt},\ and\ \citenamefont {Gerstmann}}]{Hoe11}%
  \BibitemOpen
  \bibfield  {author} {\bibinfo {author} {\bibfnamefont {F.}~\bibnamefont
  {Hoehne}}, \bibinfo {author} {\bibfnamefont {J.}~\bibnamefont {Lu}}, \bibinfo
  {author} {\bibfnamefont {A.~R.}\ \bibnamefont {Stegner}}, \bibinfo {author}
  {\bibfnamefont {M.}~\bibnamefont {Stutzmann}}, \bibinfo {author}
  {\bibfnamefont {M.~S.}\ \bibnamefont {Brandt}}, \bibinfo {author}
  {\bibfnamefont {M.}~\bibnamefont {Rohrm\"uller}}, \bibinfo {author}
  {\bibfnamefont {W.~G.}\ \bibnamefont {Schmidt}}, \ and\ \bibinfo {author}
  {\bibfnamefont {U.}~\bibnamefont {Gerstmann}},\ }\href@noop {} {\bibfield
  {journal} {\bibinfo  {journal} {Phys. Rev. Lett.}\ }\textbf {\bibinfo
  {volume} {106}},\ \bibinfo {pages} {196101} (\bibinfo {year}
  {2011})}\BibitemShut {NoStop}%
\bibitem [{\citenamefont {Press}\ \emph {et~al.}(2007)\citenamefont {Press},
  \citenamefont {Teukolsky}, \citenamefont {Vetterling},\ and\ \citenamefont
  {Flannery}}]{Pre07}%
  \BibitemOpen
  \bibfield  {author} {\bibinfo {author} {\bibfnamefont {W.~H.}\ \bibnamefont
  {Press}}, \bibinfo {author} {\bibfnamefont {S.~A.}\ \bibnamefont
  {Teukolsky}}, \bibinfo {author} {\bibfnamefont {W.~T.}\ \bibnamefont
  {Vetterling}}, \ and\ \bibinfo {author} {\bibfnamefont {B.~P.}\ \bibnamefont
  {Flannery}},\ }\href@noop {} {\emph {\bibinfo {title} {Numerical Recipes: The
  Art of Scientific Computing}}}\ (\bibinfo  {publisher} {Cambridge University
  Press},\ \bibinfo {address} {Cambridge},\ \bibinfo {year} {2007})\BibitemShut
  {NoStop}%
\bibitem [{\citenamefont {Thomas}\ \emph {et~al.}(1965)\citenamefont {Thomas},
  \citenamefont {Hopfield},\ and\ \citenamefont {Augustyniak}}]{Tho65}%
  \BibitemOpen
  \bibfield  {author} {\bibinfo {author} {\bibfnamefont {D.~G.}\ \bibnamefont
  {Thomas}}, \bibinfo {author} {\bibfnamefont {J.~J.}\ \bibnamefont
  {Hopfield}}, \ and\ \bibinfo {author} {\bibfnamefont {W.~M.}\ \bibnamefont
  {Augustyniak}},\ }\href@noop {} {\bibfield  {journal} {\bibinfo  {journal}
  {Phys. Rev.}\ }\textbf {\bibinfo {volume} {140}},\ \bibinfo {pages} {A202}
  (\bibinfo {year} {1965})}\BibitemShut {NoStop}%
\bibitem [{\citenamefont {Pierreux}\ and\ \citenamefont
  {Stesmans}(2002)}]{Pie02}%
  \BibitemOpen
  \bibfield  {author} {\bibinfo {author} {\bibfnamefont {D.}~\bibnamefont
  {Pierreux}}\ and\ \bibinfo {author} {\bibfnamefont {A.}~\bibnamefont
  {Stesmans}},\ }\href@noop {} {\bibfield  {journal} {\bibinfo  {journal}
  {Phys. Rev. B}\ }\textbf {\bibinfo {volume} {66}},\ \bibinfo {pages} {165320}
  (\bibinfo {year} {2002})}\BibitemShut {NoStop}%
\bibitem [{\citenamefont {Cohen-Tannoudji}\ \emph {et~al.}(1977)\citenamefont
  {Cohen-Tannoudji}, \citenamefont {Diu},\ and\ \citenamefont
  {Lalo\"{e}}}]{Coh77}%
  \BibitemOpen
  \bibfield  {author} {\bibinfo {author} {\bibfnamefont {C.}~\bibnamefont
  {Cohen-Tannoudji}}, \bibinfo {author} {\bibfnamefont {B.}~\bibnamefont
  {Diu}}, \ and\ \bibinfo {author} {\bibfnamefont {F.}~\bibnamefont
  {Lalo\"{e}}},\ }\href@noop {} {\emph {\bibinfo {title} {Quantum mechanics}}}\
  (\bibinfo  {publisher} {Wiley},\ \bibinfo {address} {New York},\ \bibinfo
  {year} {1977})\BibitemShut {NoStop}%
\bibitem [{\citenamefont {Akhtar}\ \emph {et~al.}(2011)\citenamefont {Akhtar},
  \citenamefont {Morishita}, \citenamefont {Sawano}, \citenamefont {Shiraki},
  \citenamefont {Vlasenko},\ and\ \citenamefont {Itoh}}]{Akh11}%
  \BibitemOpen
  \bibfield  {author} {\bibinfo {author} {\bibfnamefont {W.}~\bibnamefont
  {Akhtar}}, \bibinfo {author} {\bibfnamefont {H.}~\bibnamefont {Morishita}},
  \bibinfo {author} {\bibfnamefont {K.}~\bibnamefont {Sawano}}, \bibinfo
  {author} {\bibfnamefont {Y.}~\bibnamefont {Shiraki}}, \bibinfo {author}
  {\bibfnamefont {L.~S.}\ \bibnamefont {Vlasenko}}, \ and\ \bibinfo {author}
  {\bibfnamefont {K.~M.}\ \bibnamefont {Itoh}},\ }\href@noop {} {\bibfield
  {journal} {\bibinfo  {journal} {Phys. Rev. B}\ }\textbf {\bibinfo {volume}
  {84}},\ \bibinfo {pages} {045204} (\bibinfo {year} {2011})}\BibitemShut
  {NoStop}%
\bibitem [{\citenamefont {Paik}\ \emph {et~al.}(2010)\citenamefont {Paik},
  \citenamefont {Lee}, \citenamefont {Baker}, \citenamefont {{McCamey}},\ and\
  \citenamefont {Boehme}}]{Pai10}%
  \BibitemOpen
  \bibfield  {author} {\bibinfo {author} {\bibfnamefont {S.-Y.}\ \bibnamefont
  {Paik}}, \bibinfo {author} {\bibfnamefont {S.-Y.}\ \bibnamefont {Lee}},
  \bibinfo {author} {\bibfnamefont {W.~J.}\ \bibnamefont {Baker}}, \bibinfo
  {author} {\bibfnamefont {D.~R.}\ \bibnamefont {{McCamey}}}, \ and\ \bibinfo
  {author} {\bibfnamefont {C.}~\bibnamefont {Boehme}},\ }\href@noop {}
  {\bibfield  {journal} {\bibinfo  {journal} {Phys. Rev. B}\ }\textbf {\bibinfo
  {volume} {81}},\ \bibinfo {pages} {075214} (\bibinfo {year}
  {2010})}\BibitemShut {NoStop}%
\bibitem [{\citenamefont {{McKibbin}}\ \emph {et~al.}(2009)\citenamefont
  {{McKibbin}}, \citenamefont {Clarke}, \citenamefont {Fuhrer}, \citenamefont
  {Reusch},\ and\ \citenamefont {Simmons}}]{Mc09}%
  \BibitemOpen
  \bibfield  {author} {\bibinfo {author} {\bibfnamefont {S.~R.}\ \bibnamefont
  {{McKibbin}}}, \bibinfo {author} {\bibfnamefont {W.~R.}\ \bibnamefont
  {Clarke}}, \bibinfo {author} {\bibfnamefont {A.}~\bibnamefont {Fuhrer}},
  \bibinfo {author} {\bibfnamefont {T.~C.~G.}\ \bibnamefont {Reusch}}, \ and\
  \bibinfo {author} {\bibfnamefont {M.~Y.}\ \bibnamefont {Simmons}},\
  }\href@noop {} {\bibfield  {journal} {\bibinfo  {journal} {Appl. Phys.
  Lett.}\ }\textbf {\bibinfo {volume} {95}},\ \bibinfo {pages} {233111}
  (\bibinfo {year} {2009})}\BibitemShut {NoStop}%
\end{thebibliography}%
\end{document}